\documentclass{article}

\newcommand{\aap}{Astron. Astrophys.}

\usepackage{epsfig,rotate,multicol,graphicx,psfrag,bm,times}
\begin{document}

\title{The Neutrino Emissivity of Strange Stars with Ultra Strong Magnetic Field}
\author{Liu Xuewen, Zheng Xiaoping\thanks{email address: zhxp@phy.ccnu.edu.cn}, Hou Defu
\\
{\small  Institute of Astrophysics, Huazhong Normal University,
Wuhan430079 P. R. China }}
\date{}
\maketitle
\begin{abstract}
  The effect of a strong magnetic field on the dominant neutrino emissivity in
  strange stars is investigated. In  an ultra strong magnetic field, 
  there exists an enhanced neutrino emission because the charged particles are confined to the lowest Landau level.
  The results show that the neutrino emissivity is proportional to $T^5$ instead  of conventional $T^6$,
  which implies more rapid cooling behaviors in magnetars than usual stars\\
PACS numbers: 26.60.+c, 97.60.Jd, 12.38.Mh, 97.60.Gb
\end{abstract}
\section{INTRODUCTION}

The theoretical possibility that strange quark matter, which made
up of roughly equal numbers of up, down, and strange quarks, may
be more stable than atomic nuclei (specifically iron, which is the
most stable atomic nucleus) constitutes one of the most startling
predictions of modern physics \cite{1,2}. If Witten's idea is
correct, strange quark stars (strange stars) may exist \cite{3,4}.
Theoretical and observational researches on strange stars might
provide a scientific basis to test what is the true ground state
of the hadron.

Initially a strange star cools mainly via the neutrino emission
from its core. The $\beta$ decay of quarks in quark matter is
believed to be the important cooling mechanism in strange stars,
which was usually investigated in absence of magnetic field in
many works\cite{16,17}. However, Observations of pulsars predict
large surface magnetic field, typically of order $10^{12}G$. Some
magnetars are observed to have magnetic field of $10^{14}\sim
10^{15}G$. Considering the flux throughout the stars as a simple
trapped primordial flux, the internal magnetic field may go up to
$10^{18}G$ or even more. In spite of the fact that we do not know
yet any appropriate mechanism to produce more intense field, the
scalar virial theorem \cite{5} indeed allows the field magnitude
to be as large as $10^{20}G$.
In the case of super strong magnetic field such that\footnote{For
electrons having a Fermi momentum of 50 Mev one obtains $B>2\times
10^{17}$G, for quarks such as d quark having a Fermi momentum of
300 Mev one obtains $B<2\times 10^{19}$G.}
$2eB>p_F(e)^2$\cite{20}, all electrons  occupy the Landau
ground state with electron spins pointing in the direction
opposite to the magnetic field, and the charge neutrality now forces
the degenerate quarks also to occupy the lowest Landau level even
if the magnetic field is not large enough to break the Fermi
surface. We also know that the quantization effect on the charged
particles is extremely large when the magnetic field exceeds a
critical field strength $B_c$ defined by equating the cyclotron
energy $\frac{qB}{m_i}$ to $m_i$ (i denotes electrons and quark
flavors)\cite{19}, where the radius of cyclotron is
$\frac{1}{\sqrt{eB}}compatible to de Broglie wave length
of the particle$\cite{20}. $B_c$ will respectively reach to $\sim 10^{13},
10^{15}, 10^{18}$ gauss for electrons, u(d)-quarks and s-quarks.
Obviously, the effect on s-quark is small due to the large quark
mass if the internal magnetic field is not over $10^{18}$gauss. In
our calculations, we consider the situation that all charged
particles involved in the processes confined in the lowest Landau
level and it will give some instructions for general treatment
regardless of the internal field in stars. In this situation we
must use the exact solutions of Dirac equation.
Emission processes will be found to be evidently different  from the
field-free case. Many works recognized the magnetic effect on weak
interactions in 90's of last century \cite{6,7}. In neutron stars
the magnetic effect on the direct Urca, scattering and absorption
processes was studied extensively\cite{8,20}. We will show that
the neutrino emissivity of ultra magnetized strange star is
strongly dependent on the magnetic field.

This paper is organized as follows. The exact solution of the
Dirac equation for charged particles in the magnetic field is obtained
in Sect. 2. The direct Urca processes dominating the neutrino
emissivity in strong magnetized strange stars are calculated in
Sect. 3. The related cooling curves are presented in Sect. 4. Finally
we summarize our conclusion and discussion in Set. 5.

\section{EXACT SOLUTION OF DIRAC EQUATION FOR CHARGED PARTICLES IN MAGNETIC FIELD}
When an electromagnetic field with 4-potential $A^{\mu}$ is
included, Dirac's equation is modified by the minimal coupling
procedure of replacing $\hat{p}^{\mu}$ by
$\hat{p}^{\mu}-q\hat{A}^{\mu}$, where q is the charge of the
particle. The Dirac Hamiltonian is then replaced by
\begin{equation}
\hat{H}=\mathbf{\alpha}
\cdot(\mathbf{p}-q\mathbf{A})+\mathbf{\beta} m+q\phi
\end{equation}
Here we set $\phi=0$, just considering the charged particles in a
constant, uniform magnetic field in the z-direction. We choose the
$\emph{Landau gauge}$
\begin{equation}
\mathbf{A}=(0, Bx, 0)
\end{equation}
and assume a trial wave function of the form
\begin{equation}
\psi(t,\mathbf{x})=f(x) exp(-i \epsilon \varepsilon t + i \epsilon
p_y y + i \epsilon p_z z)
\end{equation}
where $\epsilon = \pm$ is the sign of the energy whose magnitude
is $\varepsilon$. On inserting the trial solution (3) into the
Dirac's equation in the form
\begin{equation}
(i\partial{t}-\hat{H})\psi(t,\mathbf{x)=0}
\end{equation}
one can obtain the energy and the wave function. The energy is
\begin{equation}
\varepsilon=\sqrt{p^2_z+m^2+2n|q|B}
\end{equation}
where $n$ is the Landau level. It is degenerate and can be
expressed as other quantum numbers such as the orbital part $l$
and the spin quantum number $s$.
\begin{equation}
n=l+\frac{1}{2}(1 \pm s)
\end{equation}
here the $\pm$ is on the contrary for the negative and positive
charge respectively. One has $q=|q|$ for positive charged
particles and their four solutions are
\begin{equation}
\psi^\epsilon_{+}(t,\mathbf{x})=\frac{exp[-i \epsilon \varepsilon
t + i \epsilon p_y y + i \epsilon p_z
z]}{\sqrt{2\varepsilon(\epsilon \varepsilon+m)L_y L_z}}\left(
\begin{array}{ccc}(\epsilon\varepsilon+m)I_{n;p_z}(x)\\
0\\
\epsilon p_z I_{n;p_z}(x)\\
-i \sqrt{2n|q|B} I_{n-1;p_z}(x) \end{array} \right)
\end{equation}
for spin up, and
\begin{equation}
\psi^\epsilon_{-}(t,\mathbf{x})=\frac{exp[-i \epsilon \varepsilon
t + i \epsilon p_y y + i \epsilon p_z
z]}{\sqrt{2\varepsilon(\epsilon \varepsilon+m)L_y L_z}}\left(
\begin{array}{ccc}
0\\
(\epsilon\varepsilon+m)I_{n-1;p_z}(x)\\
i \sqrt{2n|q|B} I_{n;p_z}(x)\\
-\epsilon p_z I_{n-1;p_z}(x)\\
\end{array} \right)
\end{equation}
for spin down cases. In Eqs. (7) and (8), $L_y$ and $L_z$ are the
length scales along the y and z directions. For negative particles
$q=-|q|$, the wave functions are somewhat different as
\begin{equation}
\psi^\epsilon_{q+}(t,\mathbf{x})=\frac{exp[-i \epsilon \varepsilon
t + i \epsilon p_y y + i \epsilon p_z
z]}{\sqrt{2\varepsilon_q(\epsilon \varepsilon_q+m)L_y L_z}}\left(
\begin{array}{ccc}(\epsilon\varepsilon+m)I_{n-1;p_z}(x)\\
0\\
\epsilon p_z I_{n-1;p_z}(x)\\
i \sqrt{2n|q|B} I_{n;p_z}(x) \end{array} \right)
\end{equation}
\begin{equation}
\psi^\epsilon_{q-}(t,\mathbf{x})=\frac{exp[-i
\epsilon \varepsilon t + i \epsilon p_y y + i \epsilon p_z
z]}{\sqrt{2\varepsilon_q(\epsilon \varepsilon_q+m)L_y L_z}}\left(
\begin{array}{ccc}
0\\
(\epsilon\varepsilon+m)I_{n;p_z}(x)\\
-i \sqrt{2n|q|B} I_{n-1;p_z}(x)\\
-\epsilon p_z I_{n;p_z}(x)\\
\end{array} \right)
\end{equation}
in above four Eqs. (7)-(10)
\begin{equation}
I_{n;p_z}=\left( \frac{|q|B}{\pi}
\right)^{\frac{1}{4}}exp[-\frac{1}{2}|q|B(x-\frac{p_y}{qB})^2]\times
\frac{1}{\sqrt{2^n n!}}H_n[\sqrt{|q|B}(x-\frac{p_y}{qB})]
\end{equation}
where $H_n$ is the well-know Hermite polynomial. It is convenient
to write
\begin{equation}
\xi=(\mid q\mid B)^{1/2}(x-\frac{p_y}{qB})
\end{equation}
then Eq. (11) can be rewritten as
\begin{equation}
I_{n;p_z}=\left( \frac{|q|B}{\pi}
\right)^{\frac{1}{4}}exp(-\frac{\xi^2}{2})\times
\frac{1}{\sqrt{2^n n!}}H_n(\xi)
\end{equation}

\section{NEUTRINO EMISSIVITY}
The following weak processes are important in quark matter
consisting of up, down and strange quarks,
\begin{equation}
d\rightarrow u + e^- + \bar{\nu}_e, \hspace{0.5cm} u + e^-
\rightarrow d + \nu_e.
\end{equation}
\begin{equation}
s\rightarrow u + e^- + \bar{\nu}_e, \hspace{0.5cm} u + e^-
\rightarrow s + \nu_e.
\end{equation}

We will employ the Weinberg-Salam theory to calculate the neutrino
emissivity of these reactions in strong magnetic field. The
interaction Lagrangian density for the charged current reaction
(14) and (15) may be expressed as
$\mathcal{L}_{int}=(G_f/\sqrt{2})cos\theta_c l_\mu j^\mu _W$,
where $G_f\simeq 1.435\times 10^{-49}\hspace{0.1cm} erg
\hspace{0.1cm}cm^{-3}$ is the Fermi weak coupling constant and
$\theta_c$ the Cabibbo angle. The lepton and nucleon charged weak
currents are
$l_\mu=\overline{\psi}_e\gamma_\mu(1-\gamma_5)\psi_\nu$ and $j^\mu
_W=\overline{\psi}_u\gamma^\mu(1-\gamma_5)\psi_d$ respectively.

when the u, d quarks and electrons are Landau quantized. The
emissivity due to the antineutrino emission process in the
presence of a uniform magnetic field B along z-axis is given by
\begin{eqnarray}
\dot{\varepsilon}_\nu(B)&=&
3\int_{-|q|BL_x/2}^{|q|BL_x/2}\frac{L_ydp_{dy}}{2\pi}
\int_{-\infty}^{\infty}\frac{L_zdp_{dz}}{2\pi}
\int_{-|q|BL_x/2}^{|q|BL_x/2}\frac{L_ydp_{ey}}{2\pi}\nonumber\\
& & \int_{-\infty}^{\infty}\frac{L_zdp_{uz}}{2\pi}{}
\int_{-|q|BL_x/2}^{|q|BL_x/2}\frac{L_ydp_{ey}}{2\pi}
\int_{-\infty}^{\infty}\frac{L_zdp_{ez}}{2\pi}
\int_{\infty}^{\infty}\frac{Vd^3\vec{p}_\nu}{(2\pi)^3}\nonumber\\
& & \times \varepsilon_\nu W_{fi}
f(\varepsilon_{d})(1-f(\varepsilon_{e}))(1-f(\varepsilon_{u}))
\end{eqnarray}
here we consider all charged particles are in the lowest Landau
level. The prefactor 3 takes into account the d quark's flavour
degenerate. The function $f(\varepsilon_i)$ denotes the
Fermi-Dirac functions for the $i$th particle. The transition rate
per unit volume due to the antineutrino process may be derived
from Fermi's golden rule and is given by $W_{fi}={\langle \mid
\mathcal{S}_{fi}\mid ^2 \rangle}/{tV}$. Here $t$ represents time
and $V=L_x L_y L_z$ the normalization volume. The $\mathcal{S}$
matrix element is given by
\begin{equation}
\mathcal{S}_{fi}=\frac{G_F}{\sqrt{2}}\int d^4X \overline{\psi}_e
\gamma_{\mu} (1-\gamma_5) \psi_{\overline{\nu}_e}
\overline{\psi}_u \gamma^{\mu} (1-\gamma_5) \psi_d
\end{equation}
when all the charged particles in the lowest Landau level ($n=0$),
their wave function can be expressed very simply as
\begin{equation}
\psi_e(\mathbf{x})=(\frac{eB}{\pi})^{1/4} \frac{1}{\sqrt{L_y
L_z}}exp[-i \epsilon \varepsilon_e t + i \epsilon p_{ey} y + i
\epsilon p_{ez} z] exp(-\xi_e^2)U_{e,-1}
\end{equation}
\begin{equation}
\psi_d(\mathbf{x})=(\frac{eB}{3\pi})^{1/4} \frac{1}{\sqrt{L_y
L_z}}exp[-i \epsilon \varepsilon_d t + i \epsilon p_{dy} y + i
\epsilon p_{dz} z] exp(-\xi_d^2)U_{d,-1}
\end{equation}
\begin{equation}
\psi_u(\mathbf{x})=(\frac{2eB}{3\pi})^{1/4} \frac{1}{\sqrt{L_y
L_z}}exp[-i \epsilon \varepsilon_d t + i \epsilon p_{dy} y + i
\epsilon p_{dz} z] exp(-\xi_u^2)U_{u,+1}
\end{equation}
\begin{equation}
\psi_{\nu,+1}(\mathbf{x})=\frac{1}{\sqrt{L_x L_y L_z}}exp[-i
\epsilon \varepsilon_\nu t + i \epsilon p_{\nu x} x +i \epsilon
p_{\nu y} y + i \epsilon p_{\nu z} z] U_{\nu,+1}
\end{equation}
\begin{equation}
\psi_{\nu,-1}(\mathbf{x})=\frac{1}{\sqrt{L_x L_y L_z}}exp[-i
\epsilon \varepsilon_\nu t + i \epsilon p_{\nu x} x +i \epsilon
p_{\nu y} y + i \epsilon p_{\nu z} z] U_{\nu,-1}
\end{equation}
where $\xi_e=(eB)^{1/2}(x+\frac{p_ey}{eB})$,
$\xi_d=(\frac{eB}{3})^{1/2}(x+\frac{3p_dy}{eB})$ and
$\xi_u=(\frac{2eB}{3})^{1/2}(x-\frac{3p_uy}{2eB})$. The negative
charged particles in the lowest Landau level spin down, while the
positive charged particles spin up. So from Eqs. (7)-(10), we know
that
 \begin{equation}
 U_{e,-1}=\frac{1}{\sqrt{2\epsilon \varepsilon_e(\epsilon \varepsilon_e+m_e)}}\left(\begin{array}{ccc}0\\
\epsilon\varepsilon_e+m_e\\
0\\
-\epsilon p_{ez}\\
\end{array} \right)
\end{equation}
\begin{equation}
U_{d,-1}=\frac{1}{\sqrt{2\epsilon \varepsilon_d(\epsilon
\varepsilon_d+m_d)}}
\left(\begin{array}{ccc}0\\
\epsilon\varepsilon_d+m_d\\
0\\
-\epsilon p_{dz}\\
\end{array} \right)
 \end{equation}
\begin{equation}
 U_{u,+1}=\frac{1}{\sqrt{2\epsilon \varepsilon_u(\epsilon \varepsilon_u+m_u)}}
 \left(\begin{array}{ccc} \epsilon\varepsilon_u+m_u\\
0\\
p_{uz}\\
0\\
\end{array} \right)
\end{equation}
\begin{equation}
 U_{\nu,+1}=\frac{1}{\sqrt{2\epsilon \varepsilon_\nu(\epsilon \varepsilon_\nu+m_\nu)}}
 \left(\begin{array}{ccc} \epsilon\varepsilon_\nu+m_\nu\\
0\\
\epsilon p_{\nu z}\\
\epsilon p_{\nu x}+i\epsilon p_{\nu y}\\
\end{array} \right)
\end{equation}
\begin{equation}
 U_{\nu,-1}=\frac{1}{\sqrt{2\epsilon \varepsilon_\nu(\epsilon \varepsilon_\nu+m_\nu)}}
 \left(\begin{array}{ccc} 0\\
\epsilon\varepsilon_\nu+m_\nu\\
\epsilon p_{\nu x}-i\epsilon p_{\nu y}\\
-\epsilon p_{\nu z}\\
\end{array} \right)
\end{equation}
Because only positive energy solution will contribute \cite{10}
($\epsilon=+1$), using these wave functions it is straightforward
to calculate the transition rate per unit volume which is given by
(for convenient, we take $p_{\nu x}\simeq p_{\nu, y}\simeq
p_{\nu,z}\simeq 0$.)
\begin{eqnarray}
W_{fi} & = & \frac{G_F^2cos^2\theta_c
\sqrt{2eB}(2\pi)^3}{6\sqrt{\pi}L_x L_y^2L_z^2}|M|^2
\delta(\varepsilon_d-\varepsilon_u-\varepsilon_e-\varepsilon_{\overline{\nu}_e})\nonumber\\
& &\delta(p_{dy}-p_{uy}-p_{ey}- p_{y\overline{\nu}_e})
\delta(p_{dz}-p_{uz}-p_{ez}-
p_{z\overline{\nu}_e}) \nonumber\\
& &
exp[-\frac{p_{uy}^2}{eB}-\frac{5p_{dy}^2}{2eB}-\frac{p_{ey}^2}{2eB}-\frac{p_{uy}p_{dy}}{eB}
-\frac{p_{uy}p_{ey}}{eB}+\frac{p_{dy}p_{ey}}{eB}]
\end{eqnarray}
here $M$ is the invariant matrix element. Squaring it and summing
over neutrino states, we get
\begin{eqnarray}
|M|^2 & = &
\frac{(\varepsilon_d+m_d+p_{dz})^2(\varepsilon_e+m_e-p_{ez})^2(\varepsilon_u+m_u+p_{uz})^2
}{4\varepsilon_d(\varepsilon_d+m_d)\varepsilon_u(\varepsilon_u+m_u)
\varepsilon_e(\varepsilon_e+m_e)}\nonumber\\
& &\times
\frac{[(\varepsilon_{\overline{\nu}_e}+m_{\overline{\nu}_e}-p_{z\overline{\nu}_e})^2+p_{x\overline{\nu}_e}^2
p_{y\overline{\nu}_e}^2]}{\varepsilon_{\overline{\nu}_e}
(\varepsilon_{\overline{\nu}_e}+m_{\overline{\nu}_e)}}
\end{eqnarray}
when the charged particle confined in the lowest Landau level, its
energy is $\varepsilon=\sqrt{p_z^2+m^2}$. So if the particle's
mass is zero, we have $\varepsilon=\pm p_z$. Here we consider
electron, u and d quarks' mass are zero and neglect the neutrino's
momentum, therefore only if
\begin{equation}
p_{dz}>0,\hspace{0.1cm}p_{ez},<0\hspace{0.1cm}p_{uz}>0
\end{equation}
the $|M|^2$ is not zero, it is
\begin{equation}
|M|^2=16
\end{equation}
and note that the range $L_x$ of $x$ can be fixed in a natural way
\cite{11} corresponding to the magnetic field $L_x=1/\sqrt{eB}$.
Considering all these factors, the neutrino emissivity can be
written as
\begin{eqnarray}
\dot{\varepsilon}_\nu(B) &=& 8\frac{G_F^2
cos\theta_c^2eB}{(2\pi)^6
\sqrt{\pi}}\int_{-\sqrt{1/3eB}/2}^{\sqrt{1/3eB}/2}dp_{dy}
\int_{-\sqrt{2/3eB}/2}^{\sqrt{2/3eB}/2}dp_{uy}
\int_{-\sqrt{1/eB}/2}^{\sqrt{1/eB}/2}dp_{ey} \nonumber\\
& &\delta({p_{dy}-p_{ey}-p_{uy}})
exp[-\frac{p_{uy}^2}{eB}-\frac{5p_{dy}^2}{2eB}-\frac{p_{ey}^2}{2eB}-\frac{p_{uy}p_{dy}}{eB}
-\frac{p_{uy}p_{ey}}{eB}   \nonumber\\
& & +\frac{p_{dy}p_{ey}}{eB}] \int_{-\infty}^{\infty}dp_{ez}
dp_{dz} dp_{uz} d^3\vec{p_{\nu
z}}\varepsilon_\nu f_d(1-f_e)(1-f_u)\nonumber\\
& &
\delta{(\varepsilon_d-\varepsilon_u-\varepsilon_e-\varepsilon_{\overline\nu_e})}
\delta{(p_{dz}-p_{uz}-p_{ez}-p_{\overline\nu_ez})}
\end{eqnarray}
Integrals over $p_{dy}$, $p_{uy}$ and $p_{ey}$ can be performed
numerically. The integrals over $p_{ez}$, $p_{dz}$ and $p_{uz}$
are converted into integrals over $d\varepsilon_e$,
$d\varepsilon_d$ and $d\varepsilon_u$ respectively. The integral
over neutrino solid angle is performed by using the $z$component
momentum conserving delta function. Finally using the standard
procedure \cite{12}, the neutrino emissivity can be cast into
\begin{eqnarray}
\dot{\varepsilon}_\nu(B) & = &
\frac{3\sqrt{2\pi}G_F^2cos\theta_c^2eB}{16(2\pi)^6}[\pi^2\zeta{(3)}+15\zeta{(5)}]T^5
\int_{-\sqrt{eB}/2}^{\sqrt{eB}/2}dp_{ey}\int_{-\sqrt{2eB/3}/2}^{\sqrt{2eB/3}/2}\nonumber\\
& & dp_{uy}
exp[-\frac{p_{uy}^2}{eB}-\frac{5(p_{ey}+p_{uy})^2}{2eB}-\frac{p_{ey}^2}{2eB}-\frac{p_{uy}(p_{ey}+p_{uy})}{eB}
\nonumber\\
& &-\frac{p_{uy}p_{ey}}{eB} +\frac{(p_{ey}+p_{uy})p_{ey}}{eB}]
\end{eqnarray}
Given the magnetic field, the integration in Eq. (33) can be
calculated numerically and it varies about linearly according to
the magnetic field. So the neutrino emissivity can be
approximately written as
\begin{equation}
\dot{\varepsilon}_\nu(B)\approx 6.03\times10^{26} eB_{17}^2T^5
\end{equation}
the neutrino emissivity of reaction (15) is similar to that if the
magnetic field strength is higher than $10^{18}$ gauss.

\section{COOLING CURVES OF STRONG MAGNETIZED STRANGE STAR}
Because of the coulomb barrier, strange stars may have a similar
thin crust to neutron stars\cite{3}. We postulate that the crust
of strange stars has the same structure with the same mass neutron
stars. In our calculations, we take $M=1.4M_\odot$ and $R=10km$,
so the corresponding central density of the the strange stars is
$\rho_c\simeq6.65\times 10^{14}g\hspace{0.1cm}cm^{-3}$. With
regard to the relationship between the interior temperature $T$
and the surface temperature $T_s$, we use the isothermal core
approximation which assumes that the internal temperature $T$ is
constant within the stellar core with $\rho_c\geq \rho_b =
10^{10}g\hspace{0.1cm}cm^{-3}$(Glen \& Suthland 1983) and apply
the result of Potekhin et al. (1997). We need discuss the heat
capacity besides neutrino emissivity for the calculation of the
cooling curves. We adopt the standard thermodynamic formula
(Lifshitz and Pitaevskii, 1980).
\begin{equation}
c_j=\frac{g_j}{(2\pi)^3}\int d^3\vec{p}_j(\varepsilon_j-\mu_j)
\frac{df_j}{dT}
\end{equation}
where $g_j$, $\varepsilon_j$ and $\vec{p}_j$ stand for the
particles' degeneracy, energy and momentum respectively, $\mu_j$
is the chemical potential, and $f_j$ is the Fermi-Dirac
distribution. In field-free case, we have for an almost ideal,
strongly degenerate ultra-relativistic gas
\begin{equation}
c_j=\frac{g_jp_F^2T}{6}
\end{equation}
In magnetic field, it reads
\begin{equation}
c_j=\frac{g_jeB}{(2\pi)^2}\int
dp_{zj}(\varepsilon_j-\mu_j)\frac{df_j}{dT}
\end{equation}
If we assume the charged particles to be in the lowest Landau
level in the strong magnetic field. We obtain
\begin{equation}
c_{quark}=\frac{g_jeBT}{6}
\end{equation}
We can now make cooling simulation which is reduced to solve the
equation of thermal balance
\begin{equation}
C\frac{dT}{dt}=-L_{\gamma}-L_{\nu}
\end{equation}
where $C$, $L_{\gamma}$ and $L_{\nu}$ denote respectively total
heat capacity, neutrino and surface luminosity, $C=\sum\int
c_jdV$, $L_\nu=\int \dot{\varepsilon}_\nu dV$, $L_{\gamma}=4\pi
R^2\sigma T_s^4$ with Stefan-Boltzmann const $\sigma$ and surface
temperature $T_s$. Here we will compare two cases of the s-quark
polarized and without polarization. which are showed in figure 1
and 2. In both the cases studied, the magnetic field results in
more faster cooling compared to the the field-free case.
\section{SUMMARY}
We have calculated the neutrino emissivity of strange stars with
Ultra strong magnetic field in some approximation and obtained an
analytic formula. Our results show that the neutrino emissivity is
proportional to $T^5$ instead of $T^6$ as in the field-free
case\cite{16}. Bandyopadhyay\cite{18} had already calculated the
neutrino emissivity of magnetized strange stars, but they
treated  d and s quarks  as free particles and they found that
the relation between the neutrino emissivity and the temperature
is the same as in the field-free case, with only a different 
coefficient. Further more, we find that the emissivity is strongly
dependent on the magnetic field but independent of the electron
fraction due to the extremely polarization. In the cases we studied, we show
that the magnetic field accelerates the cooling of the strange stars.
Finally, we emphasize that a general treatment must consider  the
contribution of different Landau levels of all involved charged
particles because of the finite field strength.  But we  expect
 the neutrino emissivity is still proportional to $T^5$, with only corrections
 to the coefficient.
course, the analytical calculation is difficult and the numerical
method should be applied. The future works will be necessary.
\section*{ACKNOWLEDGMENTS}
We appreciate the interesting discussion  with  Li Jiarong
and Chen Jisheng. This work was supported partly by NFSC under Grants 
90303007 , 10373007 and 10135030.

\begin{figure}[h]
\centerline{\epsfig{file=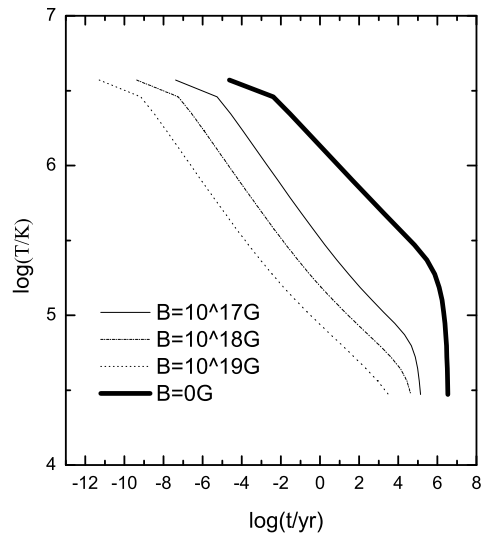, width=8cm}} \caption{Cooling
curves with the s-quark not polarized}\label{nopola}
\end{figure}
\begin{figure}[h]
\centerline{\epsfig{file=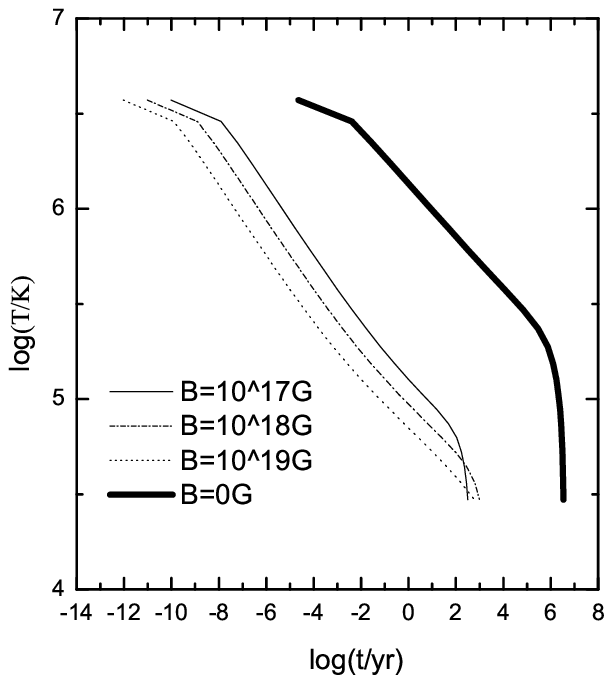, width=8cm}} \caption{Cooling
curves with the s-quark polarized}\label{pola}
\end{figure}
\end{document}